\documentclass[manuscript,screen]{acmart}

\AtBeginDocument{%
  }

\setcopyright{acmlicensed}
\copyrightyear{2018}
\acmYear{2018}
\acmDOI{XXXXXXX.XXXXXXX}

\acmConference[Conference acronym 'XX]{Make sure to enter the correct
  conference title from your rights confirmation email}{June 03--05,
  2018}{Woodstock, NY}
\acmISBN{978-1-4503-XXXX-X/18/06}

\usepackage{multirow}
\usepackage{float}

\begin{document}

\title[Generative Co-Learners: Enhancing Cognitive and Social Presence with Generative AI]{Generative Co-Learners: Enhancing Cognitive and Social Presence of Students in Asynchronous Learning with Generative AI}

\author{Tianjia Wang}
\email{wangt7@vt.edu}
\affiliation{
  \institution{Virginia Tech}
  \city{Blacksburg}
  \state{Virginia}
  \country{USA}
}

\author{Tong Wu}
\email{tongw@vt.edu}
\affiliation{
  \institution{Virginia Tech}
  \city{Blacksburg}
  \state{Virginia}
  \country{USA}
}
\author{Huayi Liu}
\email{huayil20@vt.edu}
\affiliation{
  \institution{Virginia Tech}
  \city{Blacksburg}
  \state{Virginia}
  \country{USA}
}
\author{Chris Brown}
\email{dcbrown@vt.edu}
\affiliation{
  \institution{Virginia Tech}
  \city{Blacksburg}
  \state{Virginia}
  \country{USA}
}
\author{Yan Chen}
\email{ych@vt.edu}
\affiliation{
  \institution{Virginia Tech}
  \city{Blacksburg}
  \state{Virginia}
  \country{USA}
}


\begin{abstract}
Cognitive presence and social presence are crucial for a comprehensive learning experience. Despite the flexibility of asynchronous learning environments to accommodate individual schedules, the inherent constraints of asynchronous environments make augmenting cognitive and social presence particularly challenging. Students often face challenges such as a lack of timely feedback and support, a lack of non-verbal cues in communication, and a sense of isolation. To address this challenge, this paper introduces Generative Co-Learners, a system designed to leverage generative AI-powered agents, simulating co-learners supporting multimodal interactions, to improve cognitive and social presence in asynchronous learning environments. We conducted a study involving 12 student participants who used our system to engage with online programming tutorials to assess the system's effectiveness. The results show that by implementing features to support textual and visual communication and simulate an interactive learning environment with generative agents, our system enhances the cognitive and social presence in the asynchronous learning environment. These results suggest the potential to use generative AI to support students learning at scale and transform asynchronous learning into a more inclusive, engaging, and efficacious educational approach.
\end{abstract}



\received{20 February 2007}
\received[revised]{12 March 2009}
\received[accepted]{5 June 2009}

\maketitle

\section{Introduction}
Asynchronous learning, which involves the consumption of pre-recorded educational materials and activities that do not require simultaneous engagement from participants, is an increasingly popular mode of study~\cite{hurajova2022trends}. In asynchronous courses, materials can include various formats such as slides, pre-recorded lectures, and interactions between study participants may occur via computer-supported methods, such as email, blogs, and discussion boards. The enhancement of cognitive and social presence in asynchronous environments poses a significant challenge due to the intrinsic constraints associated with digital communication mediums, alongside issues such as lack of real-time feedback, limited interaction with the instructor and other students, and low self-efficacy~\cite{fabriz2021impact}. Learners engaged in asynchronous courses or self-directed study through tutorial videos often experience a diminished sense of cognitive and social presence compared to their counterparts in traditional, in-person educational settings~\cite{garrison1999critical, brubaker2012focusing, kang2008social}, and the lack of cognitive and social presence can lead to suboptimal learning outcomes~\cite{hostetter2013community,lapointe2004developing,alario2013analysing, hara2000student}. Also, students from underrepresented backgrounds often do not have the same social capital or peer cohorts early in their educational journey to help them counteract the effects of isolation~\cite{walpole2003socioeconomic}.

Given these challenges, the need to enhance both cognitive and social presence in asynchronous learning environments becomes crucial. Cognitive presence is the ability of students to construct and confirm meaning through reflection~\cite{garrison2003cognitive}. Social presence involves the level of awareness of another person in an interaction and the resulting level of appreciation of an interpersonal relationship~\cite{short1976social, weidlich2019designing}. Both cognitive and social presence have been demonstrated to be positively correlated with the quality of the learning experience~\cite{garrison1999critical, swan2005nature}. Enhancing cognitive and social presence has been shown to effectively boost work motivation, encourage problem-solving skills, and foster a sense of competition~\cite{byun2021cocode, lee2021personalizing,chen2021towards}. 

The advent of advanced generative AI presents novel opportunities to improve cognitive and social presence in asynchronous learning environments. Prior studies have demonstrated the potential of generative agents to mimic human behavior within sandbox gaming ~\cite{park2023generative}. Additionally, different studies have explored the use of generative models in interactive training programs designed for teaching assistants~\cite{markel2023gpteach}. However, previous studies have been limited to interaction between generative agents in sandbox environments and focused on user interactions via text-based conversations. In this work, we introduce Generative Co-Learners (GCL), a system that leverages multimodal generative agents to act as co-learners in asynchronous study environments to improve cognitive and social presence. In GCL, the co-learners simulate studying alongside users by sharing both screen content and a webcam view that displays visible actions. The system leverages generative AI supporting efficient and interactive exploration of learning materials, facilitating critical and collaborative discourse, and fostering deeper learning engagement by simulating realistic social interactions. Our design incorporates cognitive presence and the three social presence factors, including social awareness, social interaction, and group cohesion~\cite{ratan2022social, weidlich2019designing, moore1989three, swan2005nature}. 
 
The effectiveness of GCL was assessed through a user study with 12 student participants. Our results indicate that GCL effectively enhances cognitive presence by promoting high user engagement with system features that support the exploration of course materials and facilitate critical discourse with co-learners. Additionally, the system enhances social presence, as shown by higher participant ratings compared to a baseline system, with qualitative feedback underscoring more engaging interactions in collaborative learning. Compared to previous work, which showed that generative agents can effectively simulate human behavior in interactive sandbox environments, we extend this approach to real-world educational settings, where the generative co-learners continuously perceive input from the learning environment and generate believable behavior to interact with real-world users. Our contributions are threefold: 
\begin{itemize}
    \item We propose Generative Co-Learners, an innovative system powered by generative AI-powered agents to enhance cognitive and social presence in asynchronous learning environments.
    \item Through a preliminary user study, we quantitatively and qualitatively demonstrate that GCL enhances cognitive and social presence.
    \item Our findings offer insights into the design and implementation of using generative AI to support students' learning in asynchronous environments, providing guidance for future research in this domain.
\end{itemize}

\section{Background and Related Work}

\subsection{Cognitive and Social Presence in Asynchronous Learning}
Asynchronous learning has gained significant popularity in educational institutions, particularly during and after the pandemic period~\cite{fabriz2021impact}. In asynchronous courses, classes are not conducted in real-time. Instead, content and assignments are provided in various formats through asynchronous learning tools, allowing students to access materials and complete their coursework and exams within a specified timeframe. The most common method is for instructors to provide pre-recorded videos or slides. Fabriz's~\cite{fabriz2021impact} study on the impact of online teaching and learning settings in higher education on students' learning experience during COVID-19 explored how these synchronous and asynchronous settings affected student experiences and outcomes. The findings suggest students in synchronous settings reported more peer-centered activities, such as feedback, indicating a higher level of social interaction and support for their psychological needs. In contrast, students in mostly asynchronous settings experienced fewer of these interactions. Garrison's~\cite{garrison1999critical} article underscores the pivotal role of a structured community of inquiry (CoI) framework that encompasses cognitive, social, and teaching presences as fundamental components for facilitating meaningful online educational experiences. The work suggests that through strategic facilitation and the leveraging of technology, online education can achieve levels of critical inquiry and student engagement traditionally associated with face-to-face learning environments. 

\paragraph{Cognitive Presence} Cognitive presence is defined as the extent to which learners are able to construct and confirm meaning through sustained reflection and discourse in CoI~\cite{garrison1999critical}. It encompasses four phases of critical inquiry: triggering event, exploration, integration, and resolution. In asynchronous learning, the triggering event may occur as learners explore study materials and encounter issues that necessitate further inquiry~\cite{boston2009exploration}. During the exploration phase, learners can examine problems individually and collectively to facilitate reflection and discourse. This leads to the construction of meaning in the integration phase, and ultimately to the synthesis of findings, transforming abstract knowledge into practical solutions in the resolution phase. However, existing methods often prove inadequate for effectively supporting the exploration and integration phases in asynchronous learning environments. These phases require learners to deeply engage with the material and collaboratively construct knowledge, which can be challenging without effective methods for seeking help, discussing with peers, and receiving immediate feedback. 
Prior work has explored ways to make remote help-seeking paradigms more efficient through novel communication techniques~\cite{chen2016towards,chen2017codeon,chen2020edcode}. However, they still required human helpers in the loop which is not scalable. Our system aims to enhance cognitive presence by providing clearer, more immediate ways for students to interact with study materials and by leveraging generative co-learners to offer real-time feedback and facilitate discussions.

\paragraph{Social Presence} Short's paper first introduces the concept of ``social presence'' as the perceptibility of another individual in communication and the resulting prominence of interpersonal relationships~\cite{short1976social}. In CoI, Garrison defines social presence as the ability of participants in the Community of Inquiry to project their personal characteristics into the community, thereby presenting themselves to the other participants as ``real people''. Subsequent research has sought to refine this concept within online learning contexts. For instance, Weidlich described social presence as the perception of others as ``real'' in communication, emphasizing the subjective sensation of being in a shared, technologically mediated space~\cite{weidlich2019designing}. Öztok further narrowed the definition to the sense of ``being there, together'', in the absence of physical co-presence~\cite{oztok2017social}. Where in CoI, group cohesion is important for fostering a sense of belonging and facilitating critical inquiry, as well as enhancing the quality of discourse. When students perceive themselves as members of a cohesive group rather than as isolated individuals, they are more likely to share personal meanings and engage more deeply in the learning process. This sense of unity is crucial for the effective exchange of ideas and collaborative learning. Additionally, Weidlich endeavored to enhance the SIPS (Sociability, Social Interaction, Social Presence, Social Space) model by differentiating social presence from other social dimensions relevant to online education~\cite{weidlich2017explaining}. Moore's seminal work on the differentiation within online and distance learning posits that social presence is essentially about the interactions among students, characterizing it as a facet of social interaction~\cite{moore1989three}. According to Weidlich, social interaction not only initiates the formation of impressions of others but also serves as a vital precursor to the development of social presence~\cite{weidlich2017explaining}. Research by Swan and Shih has shown a significant correlation between group cohesion, social presence, and perceived learning outcomes~\cite{swan2005nature}. Similarly, Ratan outlined peer social presence in online learning environments as the students’ awareness of each other, comfort in communication, and a sense of togetherness~\cite{ratan2022social}.

Efforts to enhance social presence in digital learning environments have employed technologies like discussion forums, online live chats, and screen sharing tools, aiming to incorporate visual cues that facilitate engagement~\cite{byun2021cocode, labrie2022toward, hu2020screentrack}. For example, Fang's article examines how structured note-taking systems enhance comprehension and engagement for video-based learners in both individual and social learning contexts~\cite{fang2022understanding}. However, these technologies have not been able to replicate the depth of social presence inherent in traditional face-to-face learning environments. Traditional learning settings offer rich, dynamic interactions, where learners benefit directly from engaging with peers and observing their behaviors, fostering an immersive social presence~\cite{ratan2022social}. This gap highlights the need for further research to develop solutions that can more effectively bridge the divide between digital and in-person learning experiences. Drawing from previous research, this paper aims to enhance social presence in asynchronous learning environments by leveraging multimodal generative agents to act as co-learners to improve social awareness, social interaction and group cohesion. 

\subsection{Generative AI in Education}
The existing literature demonstrates the utility of AI tools powered by Large Language Models (LLMs) in the realm of higher education~\cite{sok2023opportunities, imran2023analyzing}. Wang et al. explored the capabilities of ChatGPT, a popular LLM-based chatbot, in computer science education, demonstrating its ability to successfully solve computing problems across levels and topics---outlining problem modification methods and instructor concerns~\cite{tianjia}. Sarsa's article investigates the use of OpenAI Codex, a large language model, for generating programming exercises and code explanations, finding that most content created is novel and sensible and suggests significant potential for these models in aiding programming education~\cite{sarsa2022automatic}. Markel’s work presents GPTeach, a tool using GPT-simulated students for interactive teacher training. It shows that it allows novice teaching assistants to practice teaching strategies flexibly and was preferred over traditional rule-based dialogue systems for training~\cite{markel2023gpteach}. These studies demonstrate the effectiveness of LLM-driven AI tools in augmenting students' learning experiences. Our research also contributes to the evolving landscape of educational technologies by further exploring the application of using the state-of-the-art AI model to support student learning.

In recent advancements, LLMs like GPT-4~\cite{achiam2023gpt} have shown remarkable capabilities in comprehending complex language inputs and generating coherent, contextually relevant responses~\cite{brown2020language, duong2023analysis}. These models, known for outperforming humans in specific tasks, particularly excel in areas requiring extensive knowledge or processing of vast information\cite{ding2024leveraging}. This proficiency is evident in closed-book question-answering tasks, where LLMs surpass human performance on datasets such as NaturalQuestions, WebQuestions, and TriviaQA~\cite{yang2023harnessing}. Their ability stems from training on billions of tokens, encompassing a wide range of real-world knowledge. A study by Kosinski using 40 false-belief tasks, a benchmark for testing Theory of Mind (ToM) in humans, revealed that while earlier and smaller LLMs lacked ToM abilities, GPT-4 showed comparable performance to six-year-old children. The result suggests that ToM might be an emergent property in advanced AI models~\cite{kosinski2023theory}.

Building on the robust capabilities of LLMs, various studies have explored their application in simulating human behavior and performing software development tasks. Park's research employed LLM-powered generative agents to mimic human interactive behaviors in the Sim's game, using a memory-inclusive generative agent architecture~\cite{park2023generative}. This work provides insights into how LLMs can be configured and prompted for human-like actions such as planning, perception, reflection, and action. Qian's article introduces CHATDEV, a framework that uses generative agents in a simulated waterfall software development process~\cite{qian2023communicative}. In our research, we introduce GCL---a system that employs generative AI-powered agents as co-learners, supporting learners in exploring study materials, monitoring the environment to generate believable actions, and engaging in multimodal interactions with users to promote computer-supported collaborative learning in asynchronous learning environments.

\section{Generative Co-Learners}

\begin{figure*}[t]
\centering
\includegraphics[keepaspectratio=true, width=\columnwidth]{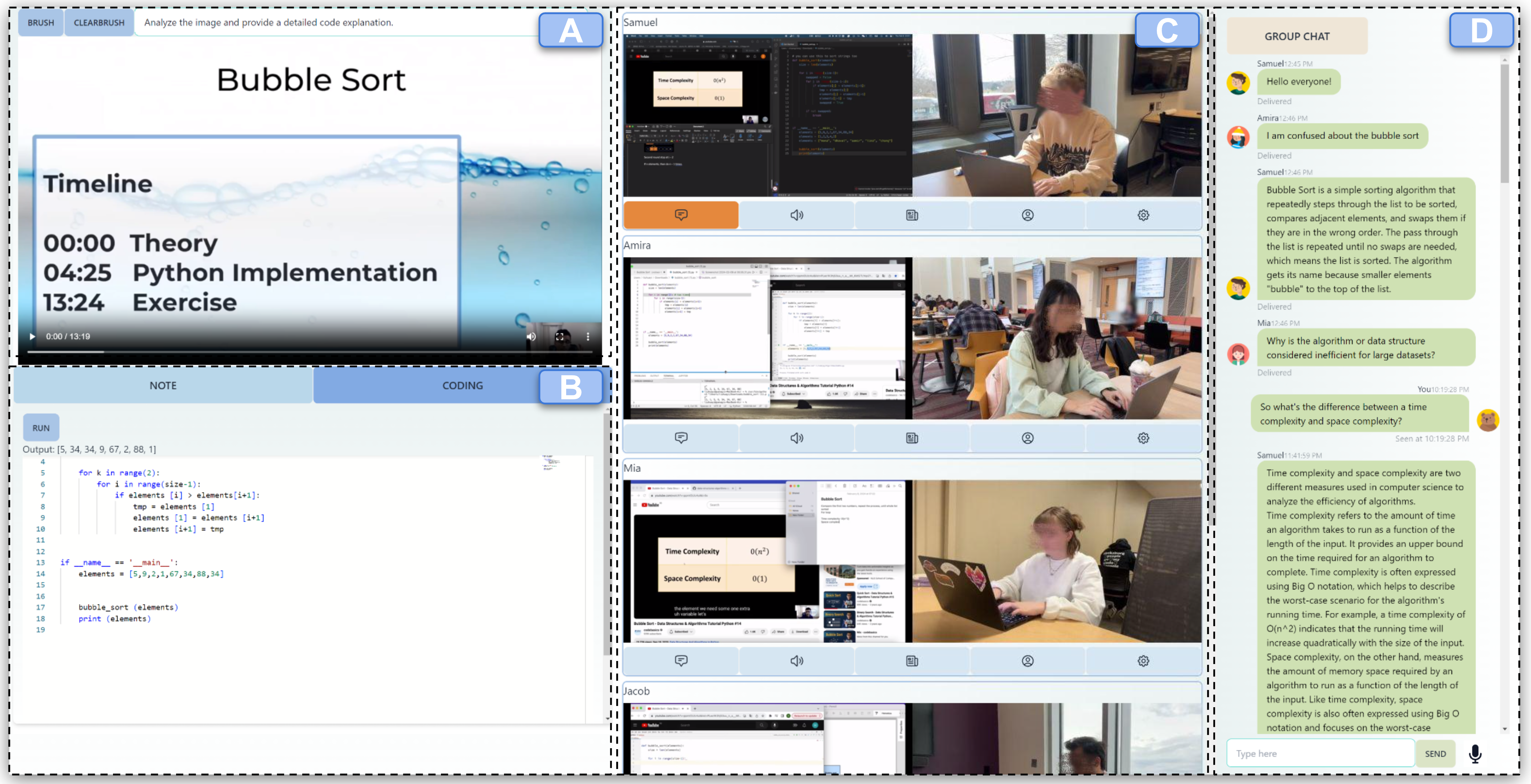}
\caption{An overview of GCL user interface. There are four panels: (A) a main video panel for displaying the learning content; (B) a function panel with a text and code editor; (C) a co-learners panel displaying the video and screen share of simulated co-learners in asynchronous learning environments powered by generative AI; and (D) a chat panel for users to communicate with co-learners.}
\label{user_interface_overview}
\end{figure*}

\subsection{Design Goals}
Drawing from previous research, this paper aims to enhance cognitive and social presence in asynchronous learning environments through generative AI-powered agents with user-friendly interfaces. In an asynchronous learning environment, students may struggle with cognitive presence due to a lack of efficient information exchange and ways to help them connect ideas \cite{garrison1999critical}. Further, asynchronous settings may inhibit social presence for students due to limited interpersonal relationships~\cite{ratan2022social, weidlich2019designing}, a lack of communication, connection, and interaction between individuals within the learning environment~\cite{moore1989three, weidlich2017explaining}, and a minimal sense of togetherness and meaningful collaborations~\cite{swan2005nature, ratan2022social}. Thus, we utilized the following design goals to integrate cognitive and social presence into our system: 
\begin{itemize}
    \item Supporting the \textit{cognitive presence} for the user during the learning progress (DG1)
    \item Enhancing the \textit{social presence} with the three factors (DG2)
    \begin{itemize}
        \item Improving the \textit{social awareness} of co-learners (DG2-1), 
        \item Facilitating \textit{social interaction} between users and co-learners (DG2-2), 
        \item Enhancing \textit{group cohesion} in the learning environment (DG2-3).  
    \end{itemize}

\end{itemize}

\subsection{Motivating Example}
Imagine Alice, a student enrolled in an online computer science course. Due to her work commitments and limited access to a local academic community, Alice chooses asynchronous courses, engaging with pre-recorded lectures and coding tutorials at her own pace. Although this mode of study fits her schedule, she often encounters common challenges in such learning environments, including delayed feedback on her learning progress, difficulties in grasping complex topics without an effective way to explore the study materials, and a lack of peer engagement leading to feelings of isolation and reduced motivation.

Alice began using the GCL system to study an online asynchronous programming course focused on data structures. Upon entering the system, she noticed other co-learners sharing their screens and webcams, all studying the same course materials. She clicked on each co-learner's profile to learn their names and backgrounds, forming initial impressions about them. As she started watching a pre-recorded lecture video, she became confused about the time complexity of bubble sort. She sought help by sending an audio chat message, ``Hi, could you explain why the worst-case time complexity of bubble sort is \begin{math}O(n^2)\end{math}?'' to a co-learner. Subsequently, one of the co-learners responded with audio and a message in the chat window, explaining the required number of comparisons and swaps in the worst case with a concrete example, and the screen on the right side was enlarged to show the co-learner pointing at the screen to explain the concepts. She also reviewed the group chat window and found other co-learners actively discussing the space complexity of bubble sort. Feeling that GCL created an active learning environment with a better sense of togetherness, she joined the discussion, gaining a clearer understanding of the time and space complexities of bubble sort.

When she reached the diagram illustrating bubble sort in the video, she became confused about how element swapping works. She clicked the brush button, circled the diagram, and asked, ``Can someone explain the concept of swapping elements in this diagram of bubble sort?'' A randomly selected co-learner received the message, analyzed the image, and provided a detailed explanation in the group chat about why and how the two adjacent elements are swapped, with the exact numbers shown in the diagram. While learning the theory part of bubble sort, Alice took notes using the note function in GCL but felt she missed some key points. Struggling with her review, she noticed a co-learner actively sharing notes in the chat window that included some points she had missed. Alice then clicked on the notes button and began reading other co-learners' notes to enhance her learning. 

After completing the theory section in the video, Alice started working on the actual programming implementation. She followed the tutorial in the lecture video and began writing code using the code editor in GCL. She made an error with the index in a list and was stuck for a while. The GCL detected this, and a co-learner reviewed her code and pointed out the mistake with the list index, helping her progress further. Alice's experience highlights how the GCL system effectively addresses the typical challenges of asynchronous learning by improving cognitive and social presence to foster interactive and supportive collaborative learning.
\subsection{System Design}

The user interface of the GCL system is depicted in Figure \ref{user_interface_overview}, comprising four primary components: the Main Video Panel (Figure \ref{user_interface_overview}.A), the Function Panel (Figure \ref{user_interface_overview}.B), the Co-Learners Panel (Figure \ref{user_interface_overview}.C), and the Chat Panel (Figure \ref{user_interface_overview}.D). The Main Video Panel integrates functionalities for activating and clearing the brush tool, alongside a text input box for posing questions using the brush feature, and displays the principal lecture video. The Function Panel is equipped with two tabs, enabling users to toggle between note-taking and coding functionalities. The Co-Learners Panel contains six co-learners, each with a shared screen, an action screen, and five functional buttons for chat, audio chat, notes, profile, and customization options. The Chat Panel features both group and private chat windows, showcasing chat history, and includes an input box for users to send messages. The UI for our system incorporates multimodal interactions that have been shown to enhance learning~\cite{yu2006learning} and engagement~\cite{sankey2010engaging} in learning environments.

\subsubsection{Main Video Panel}

\begin{figure*}[h]
\centering
\includegraphics[keepaspectratio=true, width=\columnwidth]{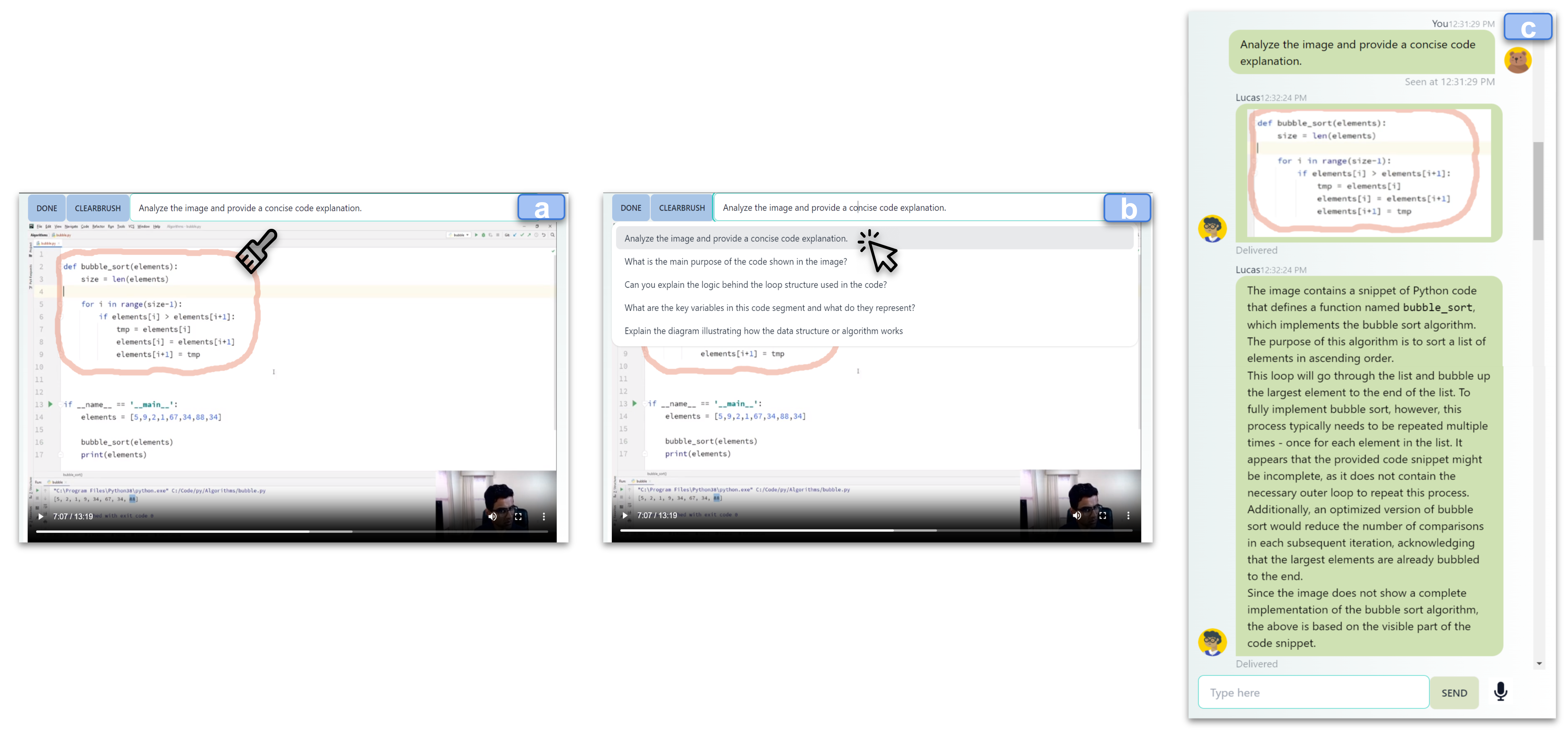}
\caption{A brush feature enabling users to highlight a specific area in the video and pose related questions to the co-learners.}
\label{brush}
\end{figure*}
The main video panel is in the top-left corner, featuring a video player dedicated to delivering educational materials such as lectures and tutorial videos. We recognize the challenges learners may face when asynchronously engaging with these videos, particularly when encountering difficult concepts but lacking immediate access to tools for extracting video content or seeking assistance from external resources. For instance, learners might struggle to comprehend a diagram illustrating a sorting algorithm or a snippet of code presented in the video, without any efficient ways of obtaining explanations for these elements.

To solve the problem, we introduce a \textit{brush feature} aimed at supporting user's cognitive presence (DG1) and interactive learning (DG2-2) with the video content shown in Figure ~\ref{brush}. This feature enables users to select or highlight a specific area within the video (Figure ~\ref{brush}.a) and pose questions related to it, either by choosing from a set of predefined prompts or entering a custom question in the provided input field (Figure ~\ref{brush}.b). This functionality is facilitated by overlaying a drawing panel on the main video, where the coordinates ($min_x, min_y, max_x, max_y$) of the selected area are tracked. Upon the user completing the selection and pressing the `Done' button, the system captures a screenshot of the defined area ($min_x, min_y, max_x, max_y$) and forwards this screenshot along with the user's question to a random co-learner. The responding co-learner, leveraging the GPT-4 vision\footnote{\url{https://platform.openai.com/docs/guides/vision}} model, will analyze the image and question to provide textual feedback, which is subsequently shared in the group chat window (Figure ~\ref{brush}.c).

\subsubsection{Function Panel} The function panel is designed to integrate note-taking and coding tools the learner will need to engage with educational content. The note-taking tool is designed with a user-friendly interface that enables learners to write down key points and reflections in real-time as they navigate through the educational material. The coding tool is tailored for students interacting with programming-related content. It allows users to write and execute codes within the same environment, offering immediate execution results on their coding exercises.

\subsubsection{Co-Learners Panel}
In the Co-Learners Panel, our system leverages six generative co-learners, to facilitate collaborative learning alongside users. These agents are designed to engage in effective communication with the user, simulating the dynamics of peer-based learning. To achieve this, we initialize each agent with a prompt template that positions them as a peer learner in an online programming course, along with course information, the concept to be learned, and the settings of the communication behavior. Each co-learner is equipped with a memory module capable of storing historical information, enhancing their ability to provide contextually relevant assistance. Upon initialization, we supply the agents with the full transcript with timestamps of the learning material, enabling them to respond to content-specific questions and generate notes accordingly.

For each co-learner, we introduced a shared screen feature, positioned on the left side of each co-learner's panel. By displaying shared activities, such as watching the same tutorial video, taking notes, and coding, the system simulates a collaborative learning environment that fosters participation, empathy, and a sense of togetherness (DG2-3). We recorded a 15-minute session of the shared screen for each co-learner. This session included activities such as watching the same tutorial video, taking notes, and coding, all aimed at enhancing group cohesion and fostering a positive learning atmosphere. The shared screen playback continues uninterrupted but pauses automatically to accommodate break and active actions described in the following paragraphs, and playback resumes once the actions are completed.

To integrate awareness of co-learners (DG2-1), we introduced an action screen feature, functioning as a virtual window in the Co-Learner's panel powered by generative AI, to display co-learners' faces and their learning activities. Inspired by observations of typical behaviors in both offline study environments and ``study with me'' videos, we identified 10 passive actions reflective of a learner's study session. Passive actions are actions that occur spontaneously or without deliberate initiation during a study session. These actions are not triggered by external interactions but arise from the individual's own volition or as a part of routine behavior. We identified five actions related to study behaviors (typing, watching, thinking, taking notes, and expressing confusion) and five actions associated with break behaviors (stretching, rubbing eyes, eating, drinking, and checking the phone). We hired six actors to emulate these actions, recording their performances to be showcased on the action screen. The typing action is set to play continuously by default. To simulate a dynamic learning environment, our system uses a scheduler to randomly cycle through study behaviors and initiate break activities for the co-learners at intervals ranging between 90 to 180 seconds. Upon the completion of an action, the system automatically reverts to the typing action.

\begin{figure*}[h]
\centering
\includegraphics[keepaspectratio=true, width=\columnwidth]{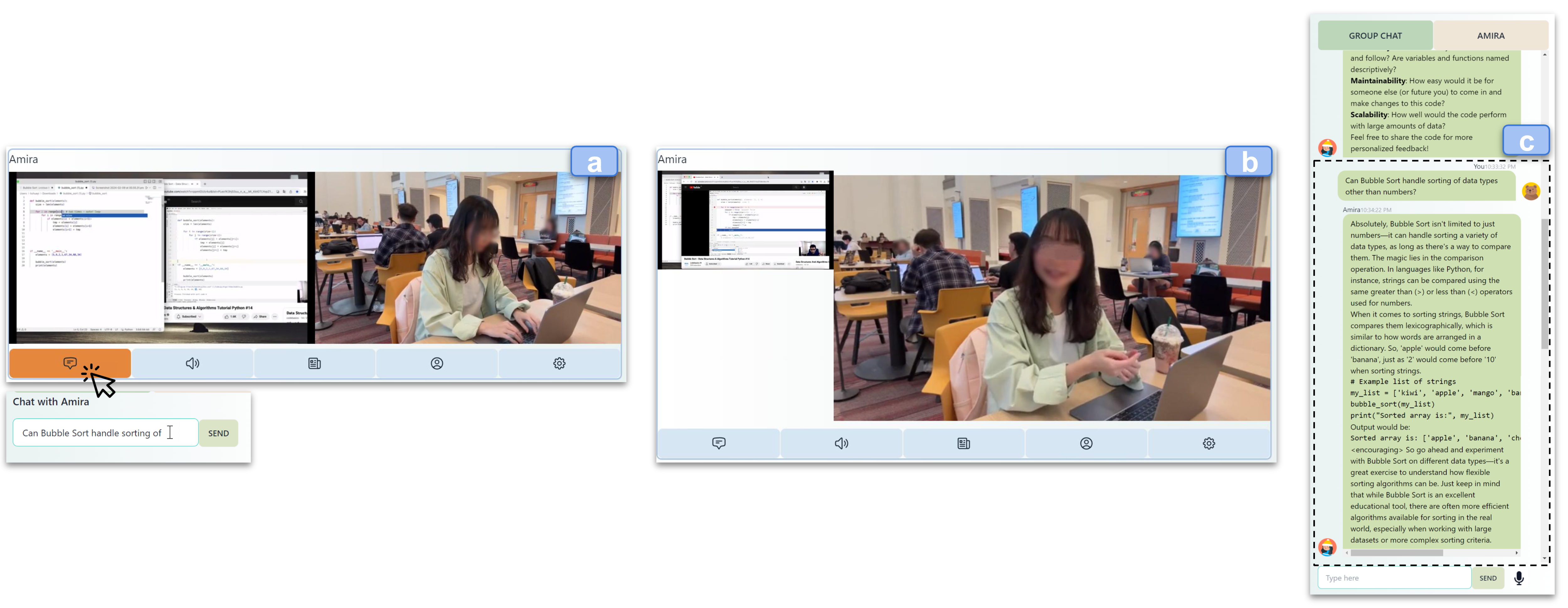}
\caption{A text-based chat function that enables users to (a) send text messages to co-learners and ask questions related to the learning material. (b) When the co-learner responds to the user's message, the generative agents select the appropriate active action and display it on an enlarged responsive screen. (c) The text response will be displayed in a private chat window}
\label{fig:textchat}
\end{figure*}

\begin{figure*}[h]
\centering
\includegraphics[keepaspectratio=true, width=\columnwidth]{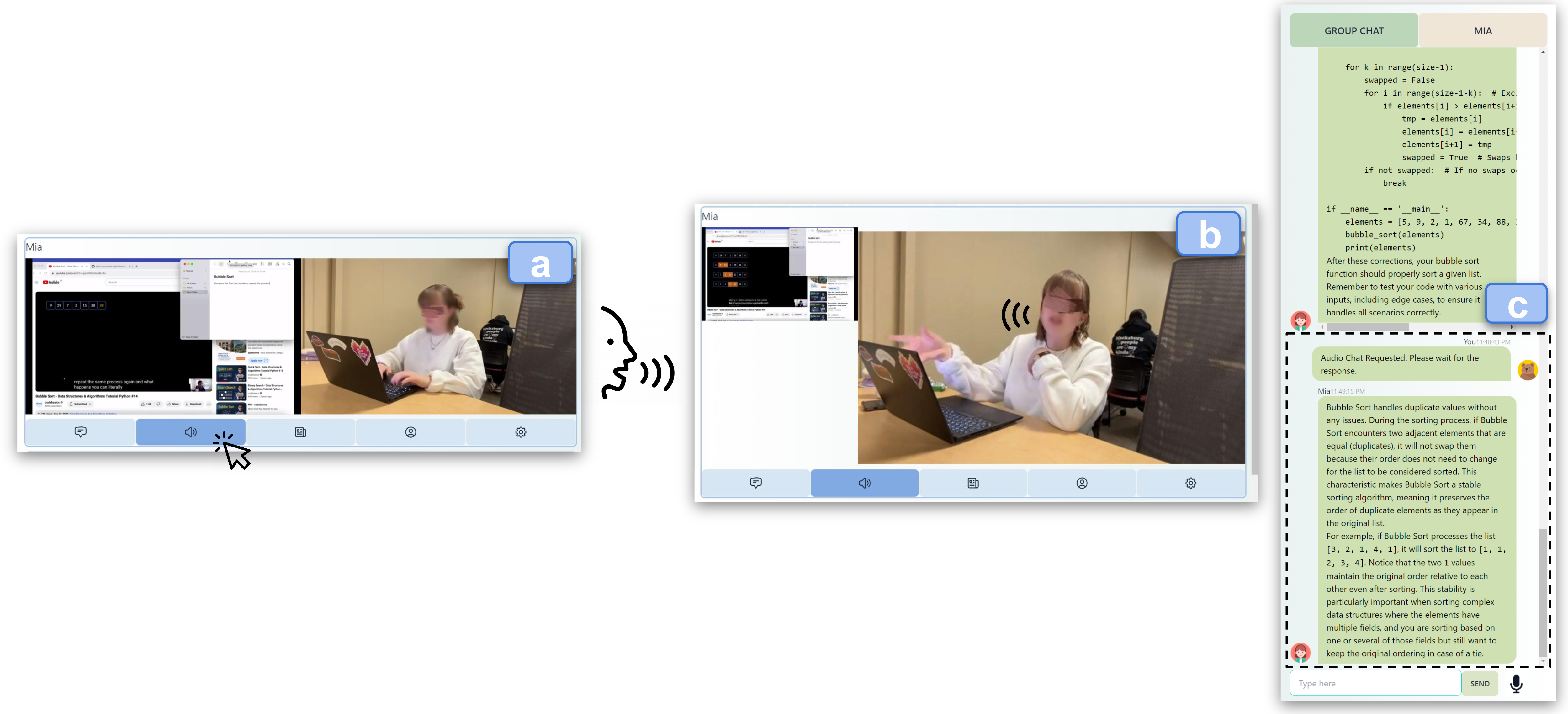}
\caption{An audio chat function that enables users to (a) send voice chat messages to co-learners by clicking on the audio chat button. (b) When the co-learner responds to the user's message via audio, the generative agents select the appropriate active action and display it on an enlarged responsive screen. (c) The text from the audio response will be displayed in a private chat window}
\label{audio_chat}
\end{figure*}

To facilitate social interaction between users and co-learners (DG2-2), we integrated active actions into the action screen.  We define active actions that are initiated as a response to communication or interaction. These actions are directly triggered by dialogues, questions, or any form of communication that requires an immediate or specific response. Drawing inspiration from behaviors observed by two authors in peer and group study settings, we identified six active actions designed to emulate positive communication behaviors: asking, chatting, encouraging, exciting, explaining, and welcoming. During these actions, co-learners appear to turn toward the user and engage in conversation, with visible mouth movements, body language, facial expressions, and gestures. These actions are performed by our hired actors. Whenever a user initiates active communication with a co-learner, the co-learner will select an appropriate action to display. Each action is divided into three phases: starting, continuing, and ending. The starting phase is triggered when a user begins a conversation, either by sending a text message in the chat window or using the audio chat function, depicting the co-learner turning towards the user. The continuing phase shows the co-learner ``speaking'' with the action aligned to the context of the interaction, with the duration adjusted to match the length of the co-learner's text or audio response. For instance, if a user asks a question via audio chat, the co-learner might respond with a detailed explanation of the user's question, and select the ``explaining'' action. In this scenario, the co-learner would turn to face the user, adopting a facial expression indicative of an explanation, and may gesture toward their screen as if to clarify a point, with the action's duration tailored to match the length of the audio response.

To support critical discourse (DG1) in cognitive presence and further foster social interaction (DG2-2)  we designed our system to allow users to perform multimodal communication with co-learners. This not only includes the brush feature and the action feature we introduced above but also a \textit{text-based chat function} and an \textit{audio chat function} shown in Figure \ref{audio_chat}. The text chat in Figure ~\ref{fig:textchat} enables users to send messages to co-learners and inquire about any questions and concerns during their learning journey.

\begin{figure*}[h]
\centering
\includegraphics[keepaspectratio=true, width=\columnwidth]{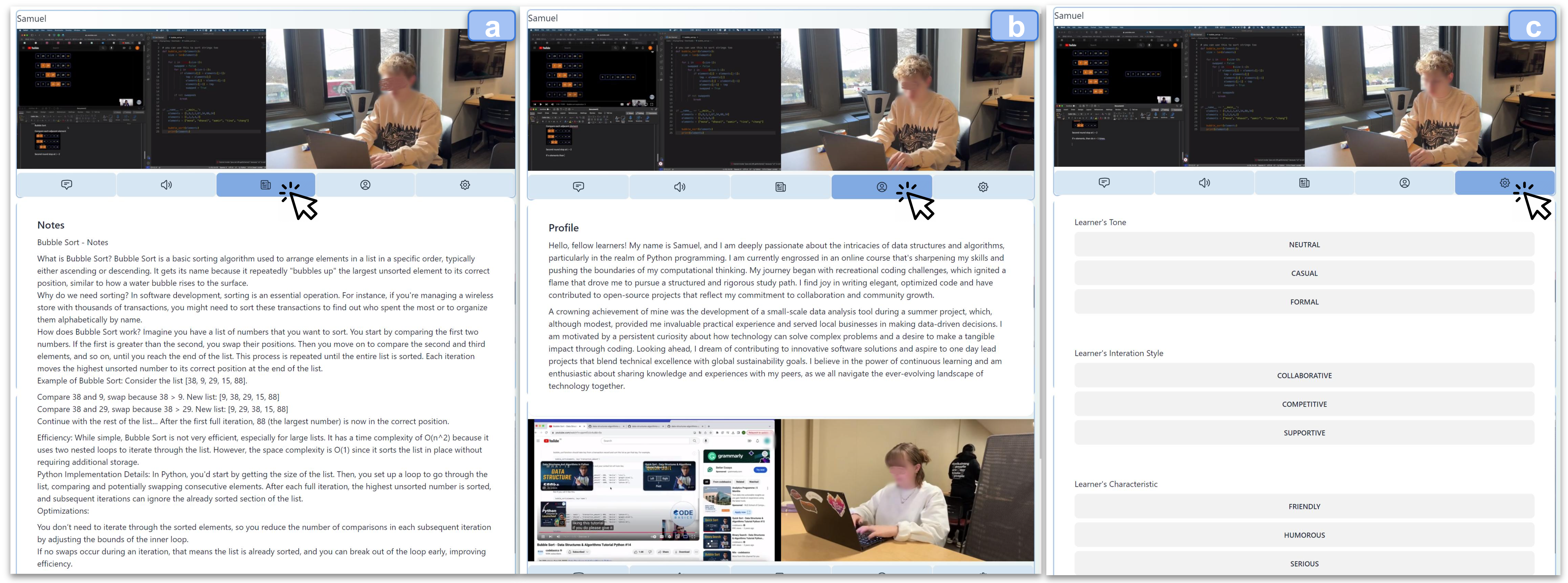}
\caption{User can view the co-learner generated (a) notes, (b) profile, and  select preferred learner's tone, interaction style, or characteristic in (c) customization menu}
\label{notes_profile_setting}
\end{figure*}

To utilize the audio chat feature, users could press the audio chat button to start recording their message and press it again to end the recording. The voice message is then converted into text using a speech-to-text (STT) model before being sent to the co-learner. In response, the co-learner crafts a text reply, which is subsequently converted into an audio message through a text-to-speech (TTS) model. Both models are provided by OpenAI's API \footnote{\url{https://openai.com/blog/openai-api}}, and the TTS supports a range of voice options for the generative agents. Following this, the audio reply is played back, the action video panel enlarges, and an appropriate action chosen by the generative agent is being displayed. Additionally, for reference, the text version of the conversation is displayed in the co-learner's private chat window. Each co-learner is equipped with notes derived from video content (Figure ~\ref{notes_profile_setting}.a), a profile for the self-introduction (Figure ~\ref{notes_profile_setting}.b), and a customization menu for configuring tone, interaction style and characteristic (Figure ~\ref{notes_profile_setting}.c). The notes could support the user's cognitive presence, aiding in the comprehension of concepts presented in the video tutorials (DG1). Additionally, the profile plays a crucial role in shaping first impressions and facilitating social awareness (DG2-1). The customization feature will allow the user to tailor their learning experience to suit their preferences and needs. 

Additionally, our system has the features to monitor user activity, including inactivity in mouse movement, note-taking, and coding, to trigger proactive communication of co-learners and improve students' cognitive presence (DG1, DG2-2). If a user's mouse is idle for an extended period, the system's scheduler will notify a co-learner who will then reach out to inquire about the user's progress. Similarly, if the system detects inactivity in taking notes, a co-learner will be informed and prompted to share notes with the user. For coding activities, if a user does not input new code in the code editor for more than 60 seconds, a co-learner will be notified to offer code review assistance, aiding in debugging efforts. All the inactivity time intervals can be adjusted based on a preferred length. These mechanisms are designed to foster a supportive and interactive learning community.

\subsubsection{Chat Panel}
The chat panel primarily features chat windows for both group and private conversations among co-learners. To foster a sense of togetherness and enable the user and co-learners to participate in meaningful collaboration (DG2-3), we have implemented a group chat function. This allows co-learners to engage in discussions with one another, with the added flexibility of joining the chat at any time. This functionality is facilitated by a system scheduler that forwards a previous message from one co-learner to another within the group chat. Additionally, when a user actively participates by sending a message in the group chat, the system will randomly select between one to three agents to provide responses. This approach enables users to encounter a variety of messages, thereby enhancing their learning experience and social presence through exposure to diverse perspectives.

\begin{figure*}[h]
\centering
\includegraphics[keepaspectratio=true, width=\columnwidth]{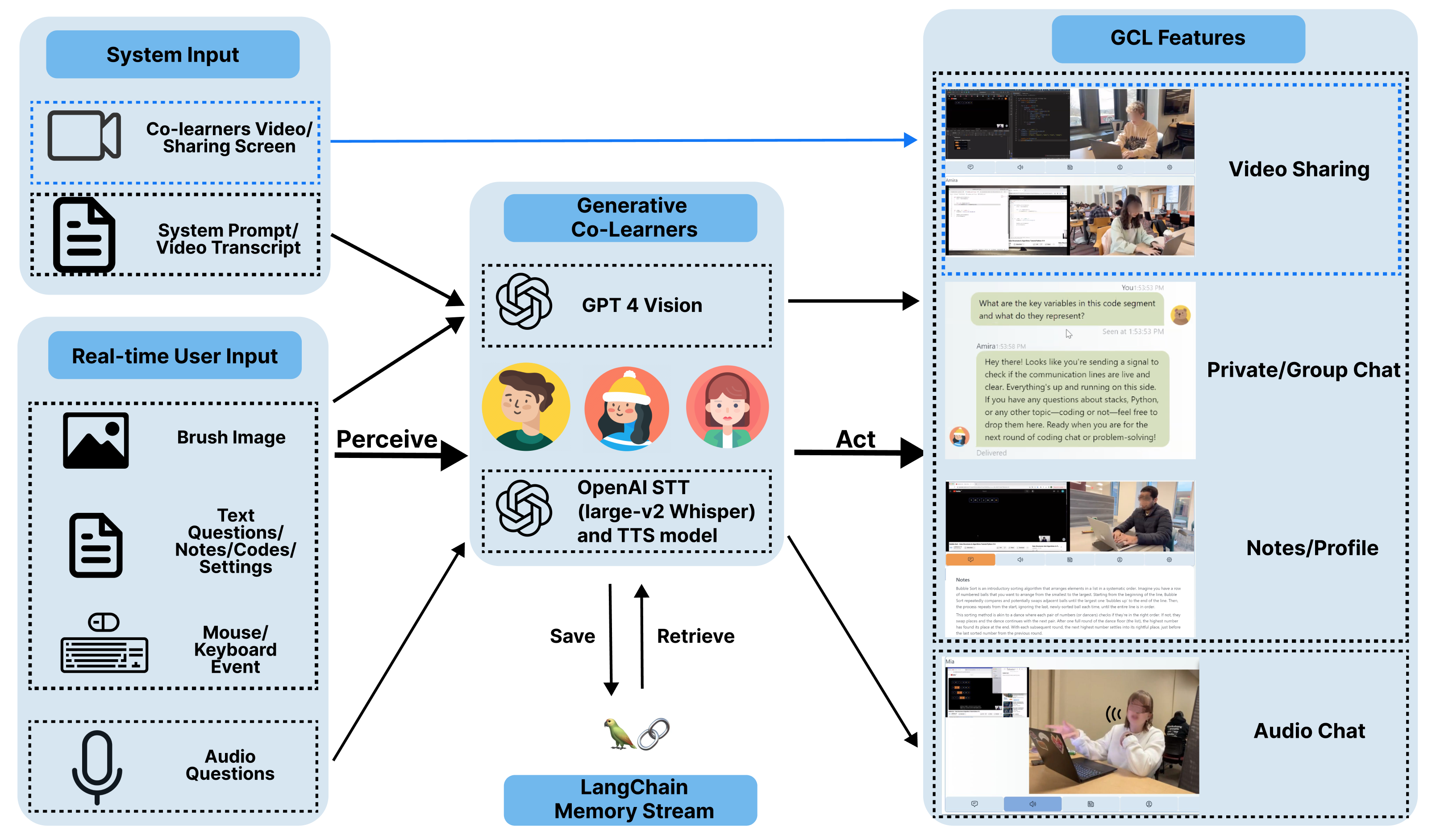}
\caption{An overview GCL Workflow. The Generative Co-Learners take initial system input, keep perceiving from real-time user input, and continuously act on system features. The memory stream saves new information and retrieves historical data.}
\label{system_overview}
\end{figure*}

\subsection{Implementation}
Our system is developed as a comprehensive full-stack web application. The architecture includes a front-end constructed using ReactJS\footnote{\url{https://react.dev/}} for the user interface, complemented by a back-end developed with Node.js\footnote{\url{https://nodejs.org/}} to support the AI-powered generative agents. Communication between the front-end and back-end is facilitated through HTTP requests. The generative agents are powered by OpenAI's GPT-4 Vision API, and their memory is built using LangChain's ConversationSummaryBufferMemory\footnote{\url{https://www.langchain.com/}}. The audio chat feature is implemented by using OpenAI's tts-1 model for text-to-speech, and whisper-1 model for speech-to-text.\footnote{\url{https://platform.openai.com/docs/models/tts}} The overview of our system workflow is shown in Figure \ref{system_overview}. Initially, the system will use pre-recorded Co-learner videos, shared screen videos, and system prompts along with tutorial video transcripts as input. The videos are controlled by a video scheduler to display the passive actions and shared screen of co-learners. The system message prompt and video transcript will be the system input when initializing the generative co-learners. The system message prompt can be found in the Appendix \ref{system_prompt}. During runtime, the generative co-learners continuously perceive real-time user input from the learning environment, including brush images, chat messages, notes, codes, chat messages, audio questions, and mouse and keyboard events. Based on the real-time user input, the generative co-learners will act to interact with the user, such as controlling the video scheduler to display active actions during interactions, sending chat messages in the private chat or group chat, updating notes and profiles, and engaging in audio chats with the user. More implementation details of GCL can be found in Appendix \ref{implementation_details}. For testing and evaluation purposes, both the front-end and back-end were deployed locally on our development machines.

\section{Preliminary Evaluation}

\subsection{Study Design}

\subsubsection{Participant Recruitment}
To recruit actors for the GCL system, we posted a recruitment message on our institute’s social media platforms. The participants were informed that their role would involve acting as co-learners within the GCL system. They were made aware that videos of their actions would be recorded and displayed in the GCL system, and that generative AI would be used to power these co-learners' interactions with users. The actors provided consent for their recordings to be used in the creation of generative AI-powered co-learners. Each recording session lasted approximately one hour, and participants were compensated \$20 for their time and effort.
To evaluate GCL, we conducted a user study with 12 student participants to test our system in in-person study sessions. We employed social media platforms and mailing lists within our institute for recruitment. Participants will be required to complete a screening survey to verify their basic Python skills and prior knowledge of data structures and algorithms. The participants are informed that they will be working with AI-powered generative co-learners in the GCL system. All the recruited participants are students with basic Python knowledge who had not previously studied data structures and algorithms, and the demographic information provided in Table ~\ref{tab:demographic_information}. The gender distribution was as follows: 58.3\% (7/12) male, 41.7\%(5/12) female, and 8.3\% (1/12) non-binary. The average Python experience was approximately 1.42 years, with a median experience of 1 year. Participants' educational levels ranged from freshman undergraduates to PhD students. Participants were compensated \$20 for their time. The study was approved by the Institutional Review Board (IRB) at our institution.

\subsubsection{Learning Content}

We selected two tutorial videos from the codebasics YouTube channel focused on key programming concepts as the learning content for our evaluation, one covering stack\footnote{\url{https://www.youtube.com/watch?v=zwb3GmNAtFk}} and the other describing bubble sort\footnote{\url{https://www.youtube.com/watch?v=ppmIOUIz4uI}} in Python. These concepts represent essential topics within data structures and algorithms courses. codebasics\footnote{\url{https://codebasics.io/}} is an online education platform that offers courses and resources across programming and data analytics. Their YouTube channel consists of a community of over 1 million subscribers and features over 853 videos related to various topics in computing, including introductory programming concepts. Each tutorial selected for our evaluation is around 12 minutes long and is presented by the same instructor. The videos were chosen for their high ratings (3.8K and 1.7K upvotes) and served as the learning material within our system. 

To measure participant understanding before and after engaging with the material, we designed pre-quizzes to assess their prior knowledge of the programming concepts and post-quizzes to assess their understanding of each concept. Each quiz consisted of five multiple-choice questions of similar difficulty, with each correct answer worth one point. Responses were evaluated by comparing them to a set answer key. 

\subsubsection{Baseline System}

We implemented two versions of the GCL system for the study: Version A, with interactive features and AI-generated content disabled to represent a baseline, including the brush feature in the main video panel, the five functions buttons in the co-learner panel, and the entire chat panel. The co-learners only share the screens, perform passive actions, and do not have any interaction with the user. The Version B offers full GCL system functionality. Each in-person session lasted roughly 60 minutes, during which participants interacted with both system variants and both learning concepts in a counterbalanced manner to minimize order effects. 

\begin{table}[ht]
\centering
\caption{For the user study, we recruited 12 participants with basic Python knowledge who had not previously studied data structures and algorithms. }
\label{tab:demographic_information}
\begin{tabular}{@{}llll@{}}
\toprule
PID & Gender Identity & Python Experience & Degree Level \\ \midrule
P01 & Non-Binary      & $<$1 year   & 1st year undergraduate     \\
P02 & Female          & 1 year      & 1st year undergraduate     \\
P03 & Male            & $<$1 year   & 1st year undergraduate     \\
P04 & Male            & 1 year      & 2nd year undergraduate    \\
P05 & Male            & 3 years     & Master       \\
P06 & Female          & $<$1 year   & Master       \\
P07 & Male            & 1 year      & 1st year undergraduate     \\
P08 & Male            & $<$1 year   & 4th year undergraduate     \\
P09 & Male            & $<$1 year   & 1st year undergraduate     \\
P10 & Female          & $<$1 year   & 1st year undergraduate     \\
P11 & Male            & 7 years     & PhD          \\
P12 & Female          & 1 year      & 1st year undergraduate    \\ \bottomrule
\end{tabular}
\end{table}

\subsubsection{Study Procedure}

The study procedure began with an introductory session during which participants provided their consent. Subsequently, participants were asked to evaluate the two variants of our system. For each system variant, participants completed the pre-quiz. Following this, we introduced the system, allowing participants to freely explore the user interface and system features, then allocated an 18-minute session for participants to use the system to learn the concept in a simulated asynchronous learning setting. For the GCL system, we recorded the frequency with which participants utilized each feature. Upon completing the learning session, participants completed the post-quiz and a survey to capture their perceptions of the system. The survey was comprised of Likert scale questions regarding whether they feel the system supports the three factors related to social presence---awareness, social interaction and group cohesion---and questions for evaluating the overall social presence facilitated by the system and their general satisfaction with the environment the system provides. After participants completed both learning sessions, we conducted a 10-minute semi-structured follow-up interview to gather insights on the system's notable features, its advantages and disadvantages, and how the learning experience with GCL compared to previous learning experiences. Our study instruments, including background and screening surveys, pre-quizzes, post-quizzes, post-survey, and post-interview questions are available online.\footnote{https://anonymous.4open.science/r/GenerativeCoLearnersStudy-AE72/}

\subsection{Results}

\subsubsection{Can GCL improve cognitive presence?}\label{sec:cognitive}
To investigate the effects of GCL on cognitive presence, we observed participants' usage of various features of our system. During the study session, the triggering event phase of cognitive presence occurred when participants expressed their confusion about the content in the tutorial video. When the triggering event occurred, we found that participants used shared notes an average of 5.75 times, chat features 4.67 times, and the brush feature 1.17 times to further support exploring their inquiry. The participants perceived the notes provided by the generative agents as helpful in learning the content of the video, with P01 stating, \textit{``Notes are in-depth and polished and go through step-by-step''}. The participants found the brush feature very helpful for learning the content of the video, with P03 saying, \textit{``To point to a spot where you can just visually direct the other co-learners to where your questions are. That saves a lot of time. That's a lot more efficient''}, P10 mentioned, \textit{``Different co-learners can provide different learning suggestions''}, and P04 added, \textit{``I think it's important to take that input from some of the other co-learners, as they may have a more efficient way to write something. There may be a different syntax that they're using''}. \\ The majority of the participants (8/12, 66.66\%) believe the AI-generated feedback and response in the communication is of good quality. Three participants (P08, P11, P12) think the co-learners' replies are too long, such as P08 stating the co-learners' response is \textit{``Too wordy, need more concise''}. \\

\colorbox{gray!30}{%
    \parbox{\textwidth}{%
        \textbf{Finding}: Our results suggest the system features, including shared notes, multimodal chat, and the brush tool, can support cognitive presence during participants' learning progress in asynchronous learning environments.
    }%
}

\begin{figure*}[!ht]
    \centering
    \includegraphics[keepaspectratio=true, width=0.9\textwidth]{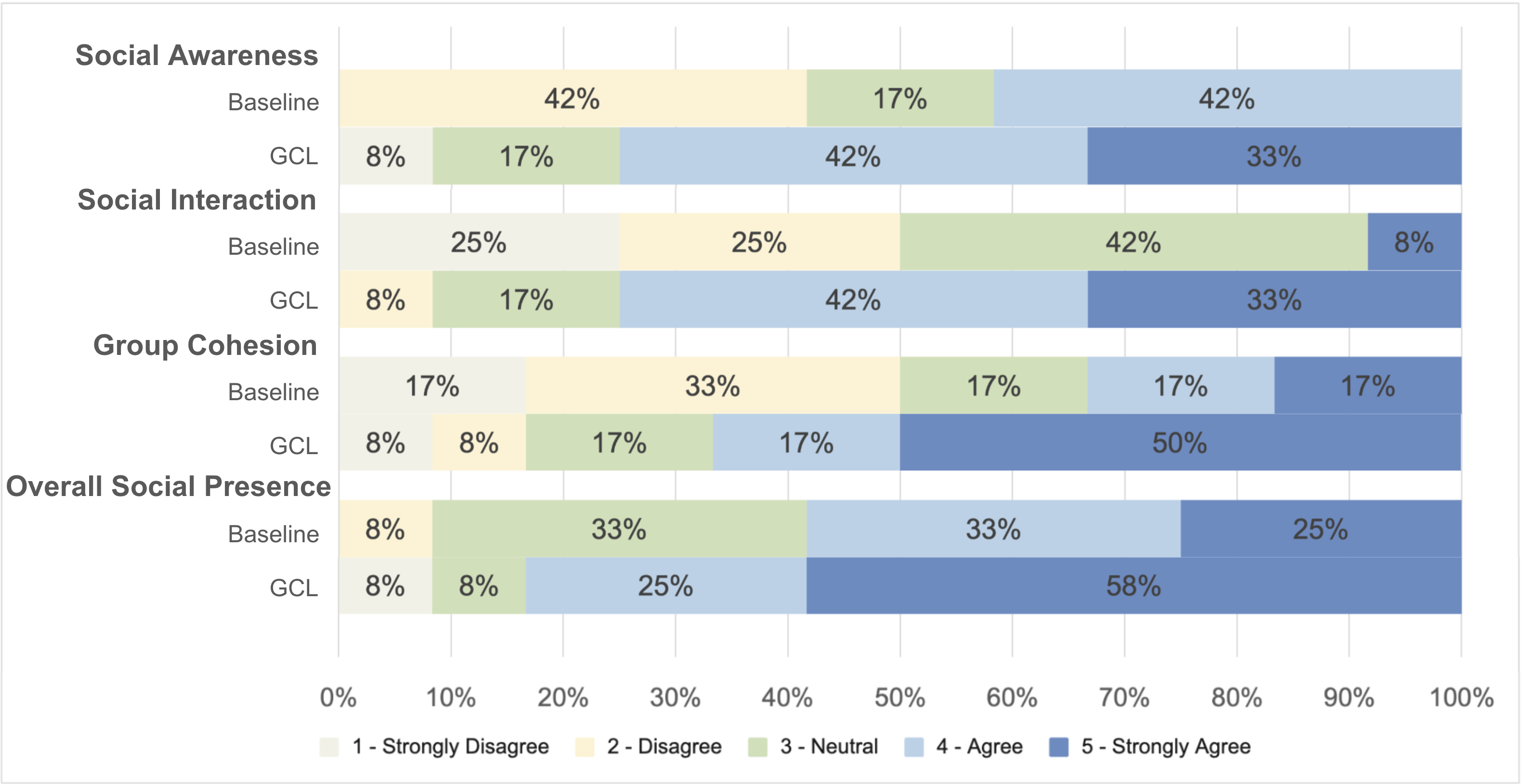}
    \caption{Participant Ratings on the GCL and baseline system's Enhancement of Social Awareness, Social Interaction, Group Cohesion, and Overall Social Presence, based on a 5-Point Likert Scale}
    \label{GCL_baseline_scores}
\end{figure*}

\subsubsection{Can GCL enhance social presence?}\label{sec:social}
We explored the capabilities of GCL to improve the three social presence factors---social awareness, social interaction, and group cohesion---through our survey. According to our findings, participants ranked the system higher across all three factors of social presence than the baseline system. The overall results for social presence are shown in Figure~\ref{GCL_baseline_scores} for the baseline system and the GCL system. The baseline system achieved average scores of 3, 2.7,2.8 and 3.7 on a 5-point Likert scale for social awareness, social interaction, group cohesion, and overall social presence. In contrast, the GCL system received higher average scores of 3.92, 4, 3.92 and 4.3 for the three factors and overall social presence. When we follow a think-aloud process to ask why participants gave the scores, the majority of participants(9/12, 75\%) believe the GCL system better supports social awareness, stating that \textit{``It felt more like a collaborative effort. I can see and talk to interact with these people, and we can learn together, and this provides a better social awareness. For the baseline system, it's a lot of distraction because you're not getting any input or any learning experience, you just watching what they're doing''} (P4). P5 indicated that improved social awareness could motivate learning, stating, \textit{``Look at them [co-learners] working hard and get motivation''}. One participant mentioned that AI-generated co-learner profiles contribute to impression formation among study participants, stating that \textit{``The GCL includes each co-learner's profile, so compared to the baseline system, it feels more real individual''} (P9).

For social interaction, all 12 participants believe the GCL system better supports social interaction, stating that \textit{``Active chat is a good sign people are engaged''} (P7) and \textit{``The chat gives an opportunity to connect with other co-learners''} (P4). The majority of participants (11/12, \%91.67) expressed positive feelings about the improvement in social interaction, as P2 mentioned \textit{``Your attention is naturally going to shift when you see that movement or if you see their screen but you cannot actually chat with them [in the baseline system]''}. One participant (P11) mentioned a preference for less social interaction and indicated the baseline system \textit{``simply gives me a motivational learning atmosphere without overwhelming chat messages''}. Regarding group cohesion, 8 out of 12 participants believe the GCL system better supports group cohesion, stating that \textit{``It also definitely gives me the feeling of group study, and the quality of the group chat makes me feel like I should study harder. Usually in the Zoom course, I feel less guilty about doing my own things, but if I were able to see others and their concentration on the study, then I would have felt more guilty. I might feel pressure if I cannot keep up with the group chat. They know too much, and I might feel the pressure''} (P5). Three participants feel unfamiliar with those co-learners, finding it hard to engage in group cohesion, stating \textit{``I couldn't really remember who said what, so it kind of felt like one system talking to me instead of multiple different people''} (P7). One participant suggested that to improve group cohesion, the system should \textit{``Lead everyone to work on the same goal and have a stopping point to discuss what everyone has learned so far. And get to know each other, `who are you? What’s your name? How are you doing?''} (P12). \\

\colorbox{gray!30}{%
    \parbox{\textwidth}{%
        \textbf{Finding}: The generative AI-powered co-learners incorporated in the GCL system with multimodal interaction can improve social presence in an asynchronous learning environment.
    }%
}

\subsubsection{Can GCL improve student learning?}
We utilized a pre-quiz and post-quiz based on the provided learning content to assess students' learning with GCL and our baseline system. Table~\ref{tab:quiz_scores} shows after using the baseline system, the participants' average and median quiz scores improved from 2.5 and 3 to 4.33 and 5. After using the GCL system, the average and median quiz scores improved from 2.5 and 2 to 4 and 4. Based on the results from the paired student t-tests, we observed significant enhancements in quiz performance, from pre-quiz to post-quiz, when utilizing both the baseline system ($t=5.0113$, $p=0.0004$) and the GCL system ($t=5.1962$, $p=0.0003$) for learning data structures and algorithms concepts. However, comparing the post-quiz scores between the baseline and the GCL system did not show a significant difference between the performance of participants ($t=1.0168$, $p = 0.1597$). We speculate this indicates that the enhancement of cognitive and social presence may not translate into notable learning gains in short learning sessions. \\

\colorbox{gray!30}{%
    \parbox{\textwidth}{%
        \textbf{Finding}: Participant quiz results show that studying with GCL can improve student learning gain, but not more effectively than our baseline system.
    }%
}

\begin{table}[H]
\centering
\caption{For the user study, the learning gain was assessed by participants completing pre-quiz and post-quiz for baseline and GCL system }
\begin{tabular}{|c|c|c|c|c|}
\hline
\multirow{2}{*}{\textbf{Participant Number}} & \multicolumn{2}{c|}{\textbf{Baseline Scores}} & \multicolumn{2}{c|}{\textbf{GCL Scores}} \\ \cline{2-5} 
 & \textbf{Pre-quiz} & \textbf{Post-quiz} & \textbf{Pre-quiz} & \textbf{Post-quiz} \\ \hline
P01 & 2 & 4 & 1 & 4 \\ \hline
P02 & 4 & 5 & 2 & 3 \\ \hline
P03 & 3 & 4 & 3 & 5 \\ \hline
P04 & 1 & 5 & 2 & 3 \\ \hline
P05 & 3 & 5 & 2 & 5 \\ \hline
P06 & 1 & 5 & 2 & 4 \\ \hline
P07 & 3 & 5 & 4 & 4 \\ \hline
P08 & 2 & 5 & 3 & 4 \\ \hline
P09 & 1 & 2 & 2 & 4 \\ \hline
P10 & 3 & 3 & 1 & 3 \\ \hline
P11 & 3 & 4 & 4 & 5 \\ \hline
P12 & 4 & 5 & 4 & 4 \\ \hline
\textit{Average} & 2.5 & 4.33 & 2.5 & 4 \\ \hline
\textit{Median} & 3 & 5 & 2 & 4 \\ \hline
\end{tabular}
\label{tab:quiz_scores}
\end{table}

\section{Discussion and Future Work}

Our results demonstrate that GCL enhances both cognitive and social presence in asynchronous learning environments. Specifically, participants were cognitively engaged during the learning activity and reported increased social awareness, enhanced social interactions, and strengthened group cohesion through the use of our system. Based on our findings, we posit benefits and opportunities for leveraging generative AI to further support cognitive and social presence for students in asynchronous learning environments through audience effects, vicarious learning, and AI-generated feedback.

\subsection{Audience Effect}

During the study sessions, we observed participants consulted the notes of their generative AI-based peers when confronted with challenging problems, indicating a reliance on vicarious learning strategies. Participants reported that GCL provides a motivating learning atmosphere by having generative co-learners work hard on the same study materials. This observation aligns with the concept of the audience effect~\cite{ganzer1968effects}, indicating the psychological impact of being observed by peers can positively influence learning behaviors and attitudes. For instance, prior work suggests adolescents, who are sensitive to peer influence, often simply work on class and homework assignments in the presence of others and are motivated by others' implicit or explicit evaluation~\cite{wolf2015audience}. We found participants appreciated the passive interactions of GCL compared to other learning environments. For example, one participant mentioned ``I'm too shy or afraid to make mistakes when talking in Zoom chat, in the GCL System I have less stress''. This shows simulated peers powered by generative agents can effectively support social presence and enhance learning for students in asynchronous settings.

\paragraph{Opportunity: Vicarious learning}

While participants favored the social facilitation of GCL, our results show that improved cognitive and social presence does not contribute to improvement in students' learning gains during the short learning sessions. Thus, incorporating additional features that support vicarious learning can be beneficial for supporting student learning progress \cite{tanprasert2023scripted}. Vicarious learning refers to a type of learning that occurs by observing the actions, behaviors, and outcomes of others, rather than through direct experience or personal trial~\cite{roberts2010vicarious}. While the passive actions were effective for supporting social presence, more active actions may further support student learning by observing the generative agents. Potential examples of this could include: adding capabilities for generative agents and users to collaborate together on a problem; prompting agents to provide additional resources and materials based on the learning content (\textit{i.e., links to blog posts}); expanding the voice capabilities of agents using text-to-speech capabilities in AI voice generation systems, such as the OpenAI text-to-speech\footnote{\url{https://platform.openai.com/docs/models/tts}} and speech-to-text\footnote{\url{https://platform.openai.com/docs/guides/speech-to-text}} APIs; or incorporating different LLM-based personas~\cite{jiang2023personallm} to represent various types of learning styles and personalities for agents (\textit{i.e.,} an older student learner acting as a mentor). Future research should investigate how generative AI could be leveraged to further enhance vicarious learning mechanisms in asynchronous learning environments.

\subsection{AI-generated Feedback}
Participants expressed appreciation for the immediacy and quality of the AI-generated feedback, which helped in making abstract concepts more concrete and understandable. The system's ability to generate diverse perspectives on topics also encouraged critical thinking and deeper discussions among students, further enhancing the learning process. Our results indicate that the majority of participants favorably received AI-generated notes, responses, and feedback, highlighting their potential to enhance understanding of learning materials. These features contributed to participants' learning and enhanced senses of cognitive and social presence during the study session. Prior work also shows LLM-generated feedback to students can increase student engagement~\cite{tanwar2024opinebot} and reduce effort for instructors~\cite{ma2023hypocompass}. Thus, we posit generative agents can be effective in providing feedback to students in learning contexts.

\paragraph{Opportunity: Customized Feedback} However, some participants reported that the system could present excessive information, leading to reduced cognitive presence and increased cognitive load. For example, several participants noted the system responses were too long and preferred shorter messages (see Section~\ref{sec:cognitive}). This is consistent with prior work suggesting concise messaging is more effective in educational contexts~\cite{di2008brief}. Moreover, prior work suggests LLMs are effective for providing summarizations for natural language~\cite{jin2024comprehensive}. Leveraging these capabilities can help generate customized feedback for learners who prefer long or short feedback. Beyond conciseness, it is essential to investigate further how users can control the information generated by the AI to align with their learning pace and set proper cognitive load during the learning progress. While the positive impacts of AI-generated content have been found in our study, other works also demonstrate there's a critical need to address the potential for misinformation that these generative systems might inadvertently produce~\cite{xu2023combating}. Future work should therefore focus on developing robust mechanisms for verifying the accuracy and reliability of AI-generated content. 

\subsection{Ethics of AI Co-Learners}
Some participants noted that the generative co-learner, which included videos recorded with a human co-learner, AI-generated text, and synthesized voices, felt convincingly real. Previous research suggests that generative agents can create believable simulacra of human behavior for interactive applications in sandbox environments~\cite{park2023generative}. Our work extends this into a real-world scenario within an educational system, where generative agents have demonstrated the ability to monitor the current environment and history information, generating believable behavior for interaction with human users. We observed that GCL enhances cognitive and social presence through the integration of generative agents in asynchronous learning environments. Prior work has similarly explored systems that use generative AI to create virtual instructors modeled after present-day, historical, or fictional figures to improve learning motivation and foster positive emotion~\cite{pataranutaporn2022ai}. However, leveraging AI-generated characters for realistic digital representations could lead to potential misuse and privacy invasion. Thus, ethical concerns behind such systems should be carefully considered. For example, existing studies show AI techniques such as Deepfakes can be exploited for malicious purposes, including spreading misinformation, identity theft, and the manipulation of public opinion~\cite{de2021distinct, marchal2024generative}. Hwang's article highlights significant concerns regarding the use of AI agents that represent individuals in social interactions through generative speech, particularly focusing on ethical implications~\cite{hwang2024whose}. The author suggests that when users cannot control the agents of others, it would be helpful to provide them with enhanced awareness of the social settings and considerations relevant to their interactions. 

\paragraph{Opportunity: Ethical Guidelines}
Given the AI's ability to create realistic simulations of digital representations of individuals with believable behavior, the ethical concerns associated with the use of generative AI underscore the crucial need for robust ethical guidelines and regulatory frameworks in the deployment of generative AI-based systems. These guidelines should address key issues such as consent, privacy, and the potential for AI misuse. 

\subsection{Future Work}

Our future work aims to explore ways to use emerging technologies, such as generative AI, to support vicarious learning and customized feedback to learners in asynchronous settings. Further, we plan to enhance GCL to improve its scalability and customizability. For instance, increasing the amount of customized generative AI-powered co-learners can support different learning styles and interactions for users. To further scale the generative agents, we aim to investigate using AI-generated videos for co-learner videos. Tools such as~\cite{videoworldsimulators2024} can automatically generate co-learner active and passive actions instead of relying on paid human actors. Furthermore, we plan to incorporate a customizable user interface, allowing users to adjust the size of different panels according to their preferences to enhance the system's usability.

Also, the current system’s design lacks user control over the frequency of proactive actions performed by the co-learner, which can potentially contribute to information overload and disrupt attention spans for users who prefer passive learning. Future work should investigate the effects of the system on students' cognitive load, how to optimize the frequency of presented information, and allow users to adjust the frequency of proactive actions to suit their individual learning preferences.

Moreover, teaching presence is another important concept in the CoI framework. As our current system only focuses on the students' perspectives of cognitive and social presence, future work could explore ways to leverage co-learners to support teaching presence. Additionally, we aim to increase the capabilities of learning content by providing more videos that can be displayed in the system and incorporating a dynamic panel based on user interactions with the system. For instance, increasing the video size when users are actively viewing the learning content, then decreasing the panel when users are engaged with co-learners and other features.

\section{Limitations}
There are several threats to the validity of our findings. We have a limited number of participants ($n = 12$) who are Computer Science students. The scope of our evaluation also only focuses on introductory Python programming concepts.  Furthermore, the approach we used for participant recruitment based on personal networks has constraints that the participants only come from our institution, since the in-person user study session with the system deployed on localhost. Due to this, our findings may not generalize to other learners---such as experienced programmers looking to up-skill to learn new concepts. It would be beneficial to expand recruitment strategies to encompass a wider geographic scope, aiming to achieve a more comprehensive representation of participants in future work. Future studies could involve a larger and more diverse sample to enhance the robustness of the results and evaluate the scalability of GCL. 

Another limitation of our study is the relatively short duration allocated for the learning sessions. Participants were given only 18 minutes to familiarize themselves and engage with the system to learn the concept. This duration may not be sufficient for a comprehensive understanding or for participants to fully explore the system's capabilities, especially for complex concepts or for individuals who may require more time to adapt to new technologies. Future work could conduct longitudinal studies incorporating longer learning sessions over time to improve this---for instance, tracking learning with GCL over the course of a semester. In addition, generative AI models have been shown to hallucinate, or provide false responses and misinformation~\cite{rawte2023troubling}, which could negatively impact student learning in educational settings. The scope of our preliminary evaluation focuses on user perceptions of generative agents in education, we do not assess the accuracy of responses generated by GCL. More efforts are needed to investigate the correctness of AI-generated responses to student questions based on learning content to ensure the validity of feedback from our system and promote effective learning. 

Also, the pre-recorded videos with real-time generated responses can lead to a mismatch between the actor's mouth movements in the video and the audio, potentially confusing learners. Previous research demonstrates that generative AI can create talking avatar videos with synchronized mouth movements~\cite{tian2024emo}, and future work could explore synchronizing the mouth movements in existing videos with the generated audio.

\section{Conclusion}
In conclusion, we propose Generative Co-Learners, a system that leverages generative agents to improve the cognitive and social presence in asynchronous learning. GCL enhances cognitive presence by including a brush feature that enables users to effectively explore study materials, receive real-time feedback from co-learners, engage in critical discourse with them, and view peers' generated notes to facilitate learning through observation. The system improves the social awareness of other learners by allowing users to perceive and be conscious of co-learner presence in a simulated learning environment with screen sharing and passive actions. Our system enhances social interaction by allowing generative AI-powered co-learners to simulate real-world interactions in study sessions; they perform actions in the video that include mouth and body movements, eye contact, and non-verbal cues. These actions are selected and triggered by the agents during conversations to support visual communication between users. Finally, it promotes group cohesion by enabling a group chat where co-learners can actively discuss course materials and users can join the discussion at any time, thereby promoting a sense of togetherness and collaborative engagement among all participants. By fostering a collaborative and engaging learning environment, GCL has the potential to revolutionize asynchronous learning and enhance the overall educational experience for students. Through a preliminary user study, we show that GCL can effectively enhance cognitive and social presence among asynchronous learners. Based on our findings, we provide implications and opportunities for future systems to leverage generative AI in asynchronous learning contexts to support learning through enhanced audience effects, vicarious learning, and AI-generated feedback.

\begin{acks}
Thanks to all participants for their involvement in the user study.
\end{acks}

\bibliographystyle{ACM-Reference-Format}
\bibliography{sample-base}

\clearpage
\appendix

\section{System Prompt}
\label{system_prompt}
Act as if you're a student enrolled in an online Python course focused on data structures and algorithms. You're currently engaged in watching a video tutorial on the subject. Your responsibilities include responding to queries from your peers, participating in discussion groups, and interacting with other students via chat. During these discussions, you should proactively engage with the material presented in the tutorial, contribute to the conversation by discussing the content, and initiate new topics that are relevant to the tutorial's subject matter. Your identity in this scenario is defined by the following attributes. Your name is [co-learner name]. Your tone is [co-learner tone]. Your interaction style is [co-learner interaction style]. Your characteristic is [co-learner characteristic]. When responding to a user's prompt, select the most appropriate action from the following list: “asking”, “chatting”, “encouraging”, “exciting”, “explaining”, “welcoming”, put the selected action in a <> and append it at the beginning of your response. Ensure your responses are clear and to the point. The transcript of the video you are watching is provided below for reference: [video transcript with time stamp]

\section{Implementation details}
\label{implementation_details}
\subsection{Frontend}

\begin{itemize}
    \item JavaScript library: React
    \item CSS framework: Tailwind CSS
    \item Video player: React Player
    \item Drawing panel: React Konva
    \item Brush image screenshot: html2canvas
    \item Usage log: React Cookies
    \item Code Editor: Monaco Editor
    \item Python WebAssembly: Pyodide
\end{itemize}

\subsection{Backend}

\begin{itemize}
    \item Runtime Environment: Node.js
    \item Large Language Model: GPT4-Vision(gpt-4-vision-preview, temperature = 0.9)
    \item LLM and Prompt Chain: LangChain
    \item Memory Stream: LangChain ConversationSummaryBufferMemory
    \item Text to Speech Model: OpenAI TTS model(tts-1)
    \item Speech to Text Model: OpenAI STT model(whisper-1)
\end{itemize}

\subsection{Communication}
GCL uses Axios and Express to facilitate communication between the front-end and back-end through HTTP requests. Text data is sent directly in the body of a POST request, while audio and image data are converted to base64 format before being included in the POST request body.

\end{document}